\documentclass[aps,pra,twocolumn,superscriptaddress,showpacs]{revtex4-1}
\usepackage{amsmath,amssymb,wasysym,graphicx}
\usepackage[utf8]{inputenc}
\usepackage[T1]{fontenc}
\usepackage{xcolor}

\IfFileExists{newtxtext.sty}
{\usepackage{newtxtext,newtxmath}}
{\IfFileExists{stix.sty}
	{\usepackage{stix}}
	{\IfFileExists{mathptmx.sty}
		{\usepackage{mathptmx}}{} } }

\usepackage{textcomp}

\usepackage{bm}

\IfFileExists{siunitx.sty}{\usepackage{booktabs,siunitx}}{}

\usepackage{color}
\definecolor{LinkColor}{rgb}{0.256,0.439,0.588}
\usepackage{hyperref}
\hypersetup{
	pdfauthor={good guys},
	pdftitle={good title},
	colorlinks=true,
	citecolor=LinkColor,
	linkcolor=LinkColor,
	urlcolor=LinkColor
}

\newcommand{\beq} {\begin{equation}}
\newcommand{\eeq} {\end{equation}}
\newcommand{\bea} {\begin{eqnarray}}
\newcommand{\eea} {\end{eqnarray}}
\newcommand{\be} {\begin{equation}}
\newcommand{\ee} {\end{equation}}

\newcommand{\ket}[1]{\left|#1\right>}
\newcommand{\bra}[1]{\left<#1\right|}

\begin{document}
\title{Dynamical phase transition and scaling in the chiral clock Potts chain}

\author{Xue-Jia Yu}
\email{xuejiayu@fzu.edu.cn}
\affiliation{Department of Physics, Fuzhou University, Fuzhou 350116, Fujian, China}
\affiliation{Fujian Key Laboratory of Quantum Information and Quantum Optics,College of Physics and Information Engineering,Fuzhou University, Fuzhou, Fujian 350108, China}

\date{\today}

\begin{abstract}
Based on time-dependent variational principle (TDVP)  algorithms, we investigate the dynamical critical behavior of quantum three-state Potts chains with chiral interactions. Using Loschmidt echo, dynamical order parameter, and entanglement entropy as an indicator, we show that as the chiral interaction $\theta$ increases, the first critical time $t_{1}^{*}$ shift towards lower values, indicating a chirality-related dynamical phase transition. Moreover, we perform dynamical scaling for the Loschmidt echo and obtain the critical exponent $\nu$ at the non-conformal critical point. The results show that as the chiral interaction $\theta$ increases, the correlation length exponent $\nu$ decreases, which is similar to the long-range interaction case. Finally, we give a simple physical argument to understand the above numerical results. This work provides a useful reference for further research on many-body physics out of equilibrium with chiral interaction.

\end{abstract}

\maketitle

\section{INTRODUCTION}
\label{sec:introduction}
Understanding exotic phases and phase transitions in many-body systems is a fundamental challenge in the field of condensed matter and statistical physics~\cite{sachdev_2011,sachdev_2023,cardy_1996,xu2012unconventional}. While extensive studies have been focused on equilibrium phase transitions~\cite{sondhi1997,fisher1972prl,fisher1974rmp,yu2022prb,yu2022pra}, less attention has been paid to the behavior of quantum many-body systems out of equilibrium~\cite{odor2004rmp,weimer2021rmp}. Dynamical quantum phase transition (DQPT) ~\cite{heyl2013prl,heyl2015prl,wdidinger2017prb,tang2023prb,Heyl_2019,Heyl_2018,zvyagin2016review} is a type of non-equilibrium phase transition that occurs at critical times $t^{*}$ during real-time evolution, characterized by non-analyticities of the rate function after a sudden quench of the system~\cite{Heyl_2018}. Analogous to equilibrium phase transitions that arise from singularities in parameter space, DQPT originates from singularities in time~\cite{heyl2013prl,Heyl_2019,Heyl_2018}. Recently, there has been a surge of interest in the study of DQPT, including investigations into critical behavior~\cite{quan2006prl,hwang2019universality,tang2023prb,Tang_2022,ding2020prb,zou2023prb,naji2022prb,jafari2022prb,naji2022pra,sadrzadeh2021prb,zamani2020prb,jadari2021pra,Mishra_2020,uhrich2020prb,Jafari2019prb,Jafari2017prl,khan2023anomalous,khan2023dynamical}, order parameters~\cite{heyl2017prb,huang2019prl,kuliashov2023prb,vosk2014prl,hagymasi2019prl,sun2020prb}, spontaneously broken symmetries~\cite{wdidinger2017prb}, and experimental realizations across a variety of platforms~\cite{tian2020prl,guopra2019,jurcevic2017prl,flaschner2018observation,wang2019prl,tian2019prb,nie2020prl,tian2020prl,wu2022dynamical}.

On the one hand, quantum phase transitions have been previously investigated to possess relativistic and conformal invariance, allowing for significant analytical progress~\cite{francesco2012conformal,ginsparg1988applied,yu2022prl}. However, there has been long debate surrounding the commensurate-incommensurate phase transition~\cite{samjdar2018pra,whitsitt2018prb,yu2022prb_fs,yu2023quantum,cheng2023prb,chepiga2019prl,maceira2022prr,Nyckees2022prr,chepiga2021prr,NYCKEES2021115365,chepiga2021kibble}, which refers to whether there is an intermediate floating phase between the commensurate-incommensurate phases studied in the 1980s, or if it is a continuous phase transition, what universality class does this phase transition belong to?  This debate has to be revisited due to the potential of neutral long-range interacting Rydberg atom array confined in optical tweezers to serve as a tunable platform for observing a variety of quantum phenomena~\cite{keesling2019quantum,serbyn2021quantum,serbyn2021quantum}.  Similarly, the $\mathbb{Z}_{3}$ clock model with chiral interaction also exhibits such unconformal chiral transition~\cite{Fendley_2012,Fendley_2014,Mong_2014,jermyn2014prb,alicea2016topological,zhuang2015prb,samjdar2018pra}. An intriguing question within the chiral clock model therefore arises: What is the relationship between the chiral and long-range interaction on quantum critical behavior?~\cite{yu2023quantum,samjdar2018pra,huang2019prb}.

On the other hand, previous studies on DQPT across different quantum critical points were carried out in many systems, such as symmetry-breaking critical point~\cite{heyl2013prl,wdidinger2017prb}, topological phase transition~\cite{budich2016prb,heyl2017prb,yang2018prb}, an exotic deconfined quantum critical point~\cite{sun2020prb}, and even non-Hermitian critical point~\cite{zhou2018pra,Tang_2022,hayata2023dissipationinduced}. Although the link between the DQPT and many physical observables has been established~\cite{Heyl_2018,Heyl_2019,zvyagin2016review}, a thorough understanding of this transition still calls for more studies. To the best of our knowledge, whether DQPT can occur in a system after a quench across a non-conformal critical point and its dynamical scaling behavior are so far less studied, therefore, it is very worthwhile to study and demonstrate the possible existence of DQPT in the system with non-conformal critical point. 

To answer the above two questions, in this work, we explore the dynamical behavior of a $\mathbb{Z}_{3}$ symmetric quantum spin chain with chiral interaction. Our investigation utilizes TDVP simulation~\cite{SCHOLLWOCK201196,schollwick2005rmp,white2004prl,yang2020prb} to examine the effect of chiral interaction, we show that the introduction of chiral interaction can advance the first critical time of DQPT. Furthermore, our analysis of the Loschmidt echo reveals that the correlation length critical exponent decreases as the chiral interaction increases, which agrees with previous studies of long-range interacting Rydberg atom arrays. The results imply chiral interaction and long-range interaction have similar effects on quantum critical behavior.

The paper is organized as follows: Section~\ref{sec:model} presents the lattice model of the quantum Potts chain with chiral interaction, the numerical method used, and the physical quantities that display DQPT. In Section~\ref{sec:potts}, we provide benchmark results for DQPT in nearest neighbor quantum Potts chains and chirality-related dynamical phase transitions. Section~\ref{sec:chiral} presents the dynamical scaling for the Loschmidt echo to obtain critical behavior in the chiral transition, as well as a simple physical explanation for our numerical observations. Finally, our conclusion is presented in Section~\ref{sec:con}. Appendixes provide additional data for our numerical calculations.

\section{MODEL AND METHOD}%
\begin{figure}[tb]
\includegraphics[width=0.6\textwidth]{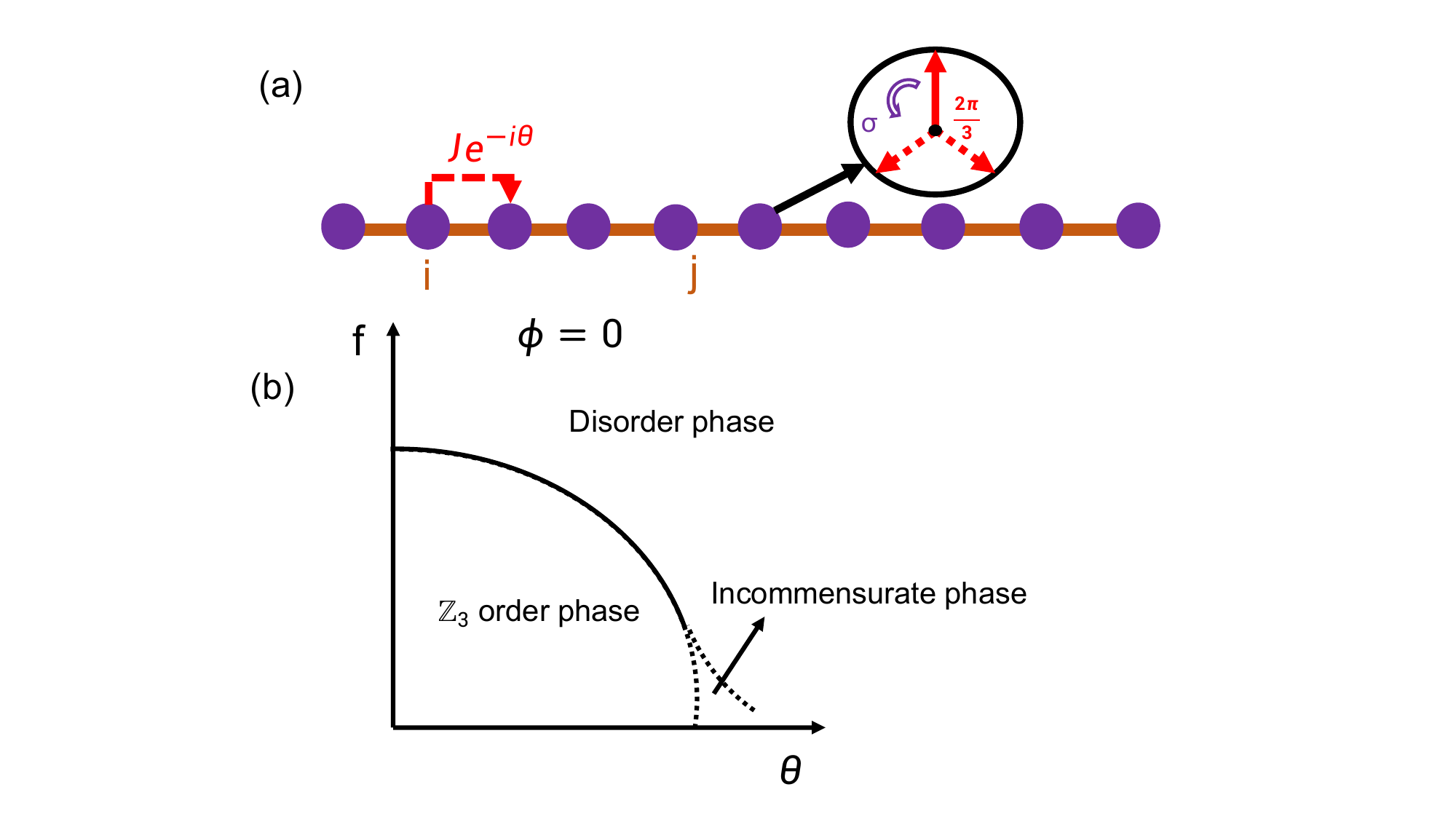}
\caption{(Color online) Schematic chiral interaction $\theta$ and $\phi$ (a) and ground phase diagram with respect to chiral interaction $\theta$ and external transverse field $f$ of quantum chiral clock Potts chain with $\phi = 0$ ~\cite{zhuang2015prb,samjdar2018pra} (b). With nonzero chiral interaction phase ($\theta \ne 0$), the effective interaction can induce incommensurate floating phases with respect to the periodicity of the underlying lattice, and transition between $\mathbb{Z}_{3}$ order phase and disorder phase belongs to non-conformal chiral universality class.}
\label{fig:phase_diagram}
\end{figure}

\label{sec:model}
The system of our study is a quantum chiral Potts chain of $L$ spins (see Fig.~\ref{fig:phase_diagram}), described by the following Hamiltonian~\cite{Fendley_2012,ostlund1981prb,huse1981prb,zhuang2015prb,samjdar2018pra,perk2016early,BAXTER1988138,AUYANG1987219}
\begin{equation}
\begin{split}
\label{E1}
&H_{CCM} = -J\sum_{j=1}^{N}\sigma_{j}^{\dagger}\sigma_{j+1}e^{-i \theta}-f\sum_{j}^{N}\tau_{j}^{\dagger} e^{-i \phi} + H.c,
\end{split}
\end{equation}
where $\phi$ and $\theta$ define two types of chiral interaction (temporal and spatial, respectively). The main text focuses on the $\phi=0$ case, where time reversal and spatial parity symmetry are both preserved but the chirality is still present as a purely spatial one. (temporal case see Appendix~\ref{sec:A2}). $J$ is the interaction strength, and $f$ represents the external transverse field. The Hilbert space is $(\mathbb{C}^{3})^{\otimes N}$. $\tau$ dictates the direction of the watch hand, and $\sigma$ rotates the watch hand clockwise through a discrete angle $2\pi/3$, as shown in Fig.~\ref{fig:phase_diagram}(a). $\sigma$ and $\tau$ satisfy $\sigma^{3}_{i}=I$,$\tau^{3}_{i}=I$, and $\sigma_{i}\tau_{j} = \omega \delta_{ij}\tau_{j}\sigma_{i}$, where $\omega = e^{2\pi i/3}$. A global $\mathbb{Z}_{3}$ transformation represented by $G = \prod_{i} \tau_{i} $ makes the Hamiltonian invariant. The operators are defined by 
\begin{equation}
\sigma=
\begin{pmatrix}
0 & 1 & 0 \\
0 & 0 & 1 \\
1 & 0 & 0
\end{pmatrix},\quad
\tau=
\begin{pmatrix}

1 & 0 & 0 \\
0 & \omega & 0 \\
0 & 0 & \omega^{2}

\end{pmatrix}.
\end{equation}

The introduction of chiral interaction has a significant impact on the phase diagram (as shown in Fig.~\ref{fig:phase_diagram}(b)) and has been extensively studied in the literature~\cite{zhuang2015prb,samjdar2018pra}. Specifically, in the absence of chiral interaction ($\theta=\phi=0$), the model reduces to the standard nearest neighbor quantum three-state Potts chain. In this case, for $f<<J$, the system is in an ordered phase that breaks the $\mathbb{Z}_{3}$ symmetry, while for $f>>J$, it is in a disordered paramagnetic phase. Fradkin-Kadanoff's transformation demonstrates that the system exhibits a continuous phase transition from the Potts-ordered topological phase to a trivial disordered phase, with a correlation length exponent of $\nu=5/6$~\cite{Fendley_2012,Fendley_2014,alicea2016topological,zhuang2015prb}. In the presence of non-zero chiral interaction, the effective interaction can be induced incommensurate floating phases relative to the lattice periodicity, and the transition between gapped states belongs to an non-conformal chiral universality class. Furthermore, the model is known to be integrable for a two-parameter family of couplings along the line $f {\rm{cos(3\phi)}}=J {\rm{cos(3\theta)}}$ and is exactly solvable.

To exhibit DQPT, we first consider the Loschmidt amplitude (return amplitude) which has been introduced by Heyl et al~\cite{heyl2013prl}
\begin{equation}
\begin{split}
\label{E2}
\mathcal{L}(t) = \langle \psi_{i}\ket{\psi(t)} = \bra{\psi_{i}} e^{-iH_{f}t}\ket{\psi_{i}},
\end{split}
\end{equation}
where $\ket{\psi_{i}}$ denotes the ground state that pertains to Hamiltonian before quench $H_{i}$ (or more generally, arbitrary initial state) and evolves under quenched Hamiltonian $H_{f}$ in time. The structure of the Loschmidt amplitude resembles the boundary partition function defined in statistical mechanics, except the time-evolution operator makes it a complex quantity instead of real. This analogy suggests the introduction of the effective free energy (return rate)
\begin{equation}
\begin{split}
\label{E3}
r(t) = -\frac{1}{N}\lim_{N \rightarrow \infty} {\rm{ln}}|\mathcal{L}|^{2},
\end{split}
\end{equation}
similar to the equilibrium statistical physics, where phase transitions are identified by singularities in the free energy at certain values of the control parameter, DQPT is characterized by non-analytic cusps in the return rate $r(t)$ at the time $t^{*}$. However, a specific non-equilibrium protocol, such as a quantum quench, can be employed to observe DQPT by driving the system out of equilibrium. In the case of a quantum quench, the initial state is prepared as the ground state of an initial Hamiltonian ($H_{0}$), and then the control parameter of the Hamiltonian is suddenly switched to a different value, leading to the final Hamiltonian ($H_{f}$).

The relationship between the dynamical order parameter and DQPT is an intriguing topic that has recently attracted considerable interest. The dynamical order parameter accurately characterize the behavior of the Loschmidt echo, specifically, the periodic behavior exhibited by the dynamical order parameters should be the same as the period of the Loschmidt echo~\cite{sun2020prb}. This definition does not necessitate conformity with the characteristics of order parameters in equilibrium quantum phase transitions.To explore this relation in the non-conformal quantum critical point, we introduce a dynamical order parameter defined as $Q(t) = \frac{1}{L}\bra{\psi(t)}\sum_{j}(\tau_{j}+\tau^{\dagger}_{j})\ket{\psi(t)}$.

Additionally, the relationship between DQPT and entanglement structures has also been investigated. DQPT may correspond to regions of rapid growth or peaks in entanglement entropy~\cite{jurcevic2017prl,markus2018scipost}. To further probe DQPT, we define the entanglement entropy as $S(t)=-{\rm{Tr}}(\rho_{A} {\rm{log}}\rho_{A})$, where $\rho_{A} = {\rm{Tr}}_{B}\ket{\psi(t)}\bra{\psi(t)}$ is the reduced density matrix about half chain A: $1,2,..., N/2$ (B: $N/2+1,..., N$).

Except for specific integrable lines, the quantum chiral clock Potts chain does not have exact solutions. In the parameter region of interest, we employ a state-of-art, time-dependent density matrix renormalization group (tDMRG) method (more precisely, TDVP method)~\cite{SCHOLLWOCK201196,schollwick2005rmp,white2004prl,yang2020prb}, based on matrix product states (MPS), which is a powerful numerical method for one-dimensional strongly correlated many-body systems. We set the bond dimension to 60, ensuring good convergence of the dynamical physical quantities (see the Appendix.~\ref{sec:A6}) by requiring relative energy errors less than $10^{-5}$. The time step during the evolution is set to $dt = 0.001$ (ensuring good convergence, see the Appendix.~\ref{sec:A1}). To minimize edge effects, we impose periodic boundary conditions and use $J=1$ as the energy unit. It's worth emphasizing that, in our cases, the finite-size TDVP algorithm is more efficient than the iDMRG algorithm for computing other physical quantities, such as entanglement entropy.

\section{Dynamical phase transition}
\label{sec:potts}
As a benchmark, we first study DQPT of the nearest neighbor quantum three-state Potts chain ($\theta=\phi=0$)~\cite{karrasch2017prb,wu2019dynamical,wu2020prb}. More precisely, we study the time evolution of the return rate after sudden quenches between the paramagnetic (PM) and Potts ferromagnetic phase (Potts FM).

We first consider a special limit in which the return rate can be obtained analytically: Start from the perfect PM phase ($f_{0} = \infty$) and quenched to the classical Potts FM phase ($f_{1} = 0$). The initial state is given by 

\begin{equation}
\begin{split}
\label{E6}
\ket{\psi_{0}} = \frac{1}{3^{N/2}}\prod_{i}(\ket{A}_{i} + \ket{B}_{i} + \ket{C}_{i}), 
\end{split}
\end{equation}
where $\ket{A}_{i}$, $\ket{B}_{i}$, and $\ket{C}_{i}$  are three degenerate Potts FM ground state, respectively. Since the final Hamiltonian is purely classical, the return amplitude can write the simple form 
\begin{equation}
\mathcal{L}(t) = \rm{tr} {\rm{M^{N}}},\quad
M=
\begin{pmatrix}
e^{2i Jt}/3 & e^{-i J t}/3 & e^{-i J t}/3 \\
e^{-i J t}/3 & e^{2 i J t}/3 & e^{-i J t}/3 \\
e^{-i J t}/3 & e^{-i Jt}/3 & e^{2 i Jt}/3
\end{pmatrix},
\end{equation}
where periodic boundary conditions on a chain with $N$ lattice sites have been considered. The eigenvalues of the transfer matrix $M$ are given by 
\begin{equation}
\begin{split}
\label{E7}
\lambda_{1} = \frac{e^{-i Jt}}{3}(e^{3 i Jt}+2), \lambda_{2} = \lambda_{3} = \frac{e^{-i Jt}}{3}(e^{3 i Jt}-1)
\end{split}
\end{equation}
and we obtain the return amplitude $\mathcal{L}(t)  = \lambda_{1}(t)^{N} + 2\lambda_{2}(t)^{N}$, which yields the return rate 
\begin{equation}
\begin{split}
\label{E8}
&l(t) = -\frac{1}{N} {\rm{ln}}|(9{\rm{cos}}^{2} \tilde{t}+{\rm{sin}}^{2}\tilde{t})^{N} +  \\
&4^{N+1}{\rm{sin}}^{2}\tilde{t}+2(2i)^{N}(3{\rm{cos \tilde{t}}}+i {\rm{sin}}\tilde{t})^{N}{\rm{sin}}^{N}\tilde{t} \\
& +2(2i)^{N}(-3{\rm{cos}}\tilde{t}+i {\rm{sin}}\tilde{t})^{N}{\rm{sin}}^{N}\tilde{t}|+2{\rm{ln}}3,
\end{split}
\end{equation}
where $\tilde{t} = 3Jt/2$. The return rate is periodic $l(t) = l(t+2\pi/3J), l(0)=0$, and shows non-analytic behavior at the critical times $Jt^{*} = 2\pi/9+2\pi n /3, n \in \mathbb{N}_{0}$, as shown in Fig.~\ref{fig:potts1}.

\subsection{Warm up: quantum Potts chains}
\begin{figure}[tb]
\includegraphics[width=0.52\textwidth]{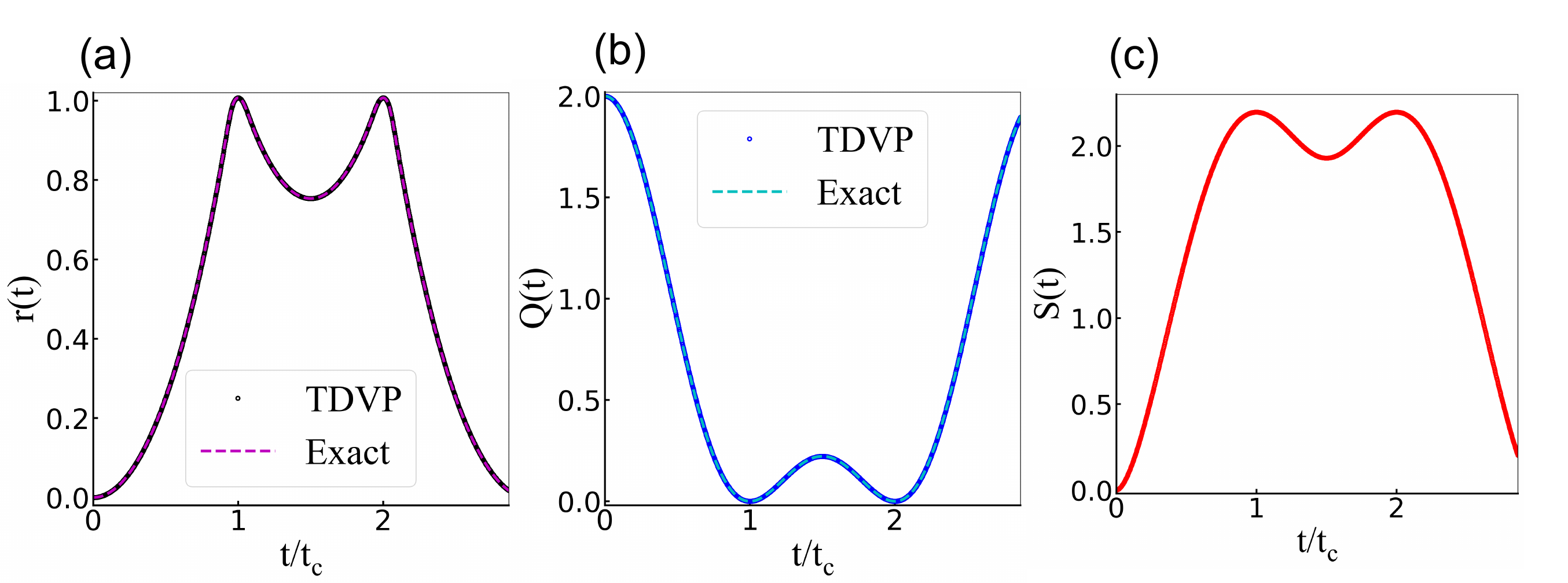}
\caption{(Color online) Time evolution of the (purple/cyan dashed line represent exact analytic results by transfer matrix) return rate (a), dynamical order parameter (b), entanglement entropy (c). All the figures are correspond to the quenches with $f_{0}/J = \infty$ (PM) $\rightarrow $ $f_{1}/J = 0.0$(Potts FM), $\theta=\phi=0, N=30$. $t_{c}=2\pi/9$ is the first critical time for the three-state Potts chain as the system size $N$ tends to $\infty$.}
\label{fig:potts1}
\end{figure}

In this section, we numerically investigate DQPT after quenches from the PM to the Potts FM phase. The quench protocol is implemented by suddenly switching the ratio between the transverse field $f$ and ferromagnetic interaction $J$ from its initial $f_{0}/J$ to its final value $f_{1}/J$. We start from the PM state, which is obtained from initial Hamiltonian $H_{0}$ with $f_{0}=1.0, J=0.0, f_{0}/J=\infty$, quenched to a final state with parameter $f_{1}=0.0, J=1.0, f_{1}/J=0.0$, located in Potts FM order phase.  The general relation between the DQPT and the underlying equilibrium quantum critical point is unclear, but as shown in previous works~\cite{Heyl_2019,Heyl_2018}, it is argued that the DQPT usually occurs when the quenching process is ramped through an equilibrium critical point. Indeed, as shown in Fig.~\ref{fig:potts1} (a), the return rate exhibits non-analytical behavior with respect to time, implying the DQPT occurs (see the Appendix~\ref{sec:A1} for the finite-size scaling), which exhibits a DQPT, and the time evolution behavior of return rate, is the same as iDMRG results.~\cite{karrasch2017prb}. Moreover, to explore the relationship between the return rate and the zeros of a dynamical order parameter, we also calculate the dynamical order parameter. During the time evolution, the dynamical order parameter exhibits periodic changes with time, with its valleys corresponding to the peaks in the return rate, as shown in Fig.~\ref{fig:potts1} (b). Our numerical observation is fully consistent with previous studies ~\cite{karrasch2017prb}. 

Moreover, we can explore the entanglement structures and possible connections to the above observations. The entanglement entropy is efficient physical quantities to uncover entanglement structures of the model. To be more precise, the half-chain entanglement entropy, are singular values of the Schmidt decomposition across a bond, is easily accessed through the finite-size DMRG calculation. In Fig.~\ref{fig:potts1} (c), sudden changes in the entanglement entropy are seen in the vicinity of the serval critical times, which is similar to the dynamical behavior of the return rate. The reason for this remains currently unclear but suggests some deeper relationship between entanglement structure and DQPT.  It is important to note that while the dynamical order parameters and entanglement entropy exhibit peaks, these peaks do not increase with system size (up to $N=30$) (see Appendix~\ref{sec:A1}). This suggests that the dynamical order parameter or entanglement entropy alone cannot be used as an indicator of whether a DQPT has occurred at this time. However, it is noteworthy that the period of change in the dynamical order parameter and entanglement entropy coincides with the period of the DQPT. 

\subsection{Chirality-related dynamical phase transition}
\begin{figure}[tb]
\includegraphics[width=0.52\textwidth]{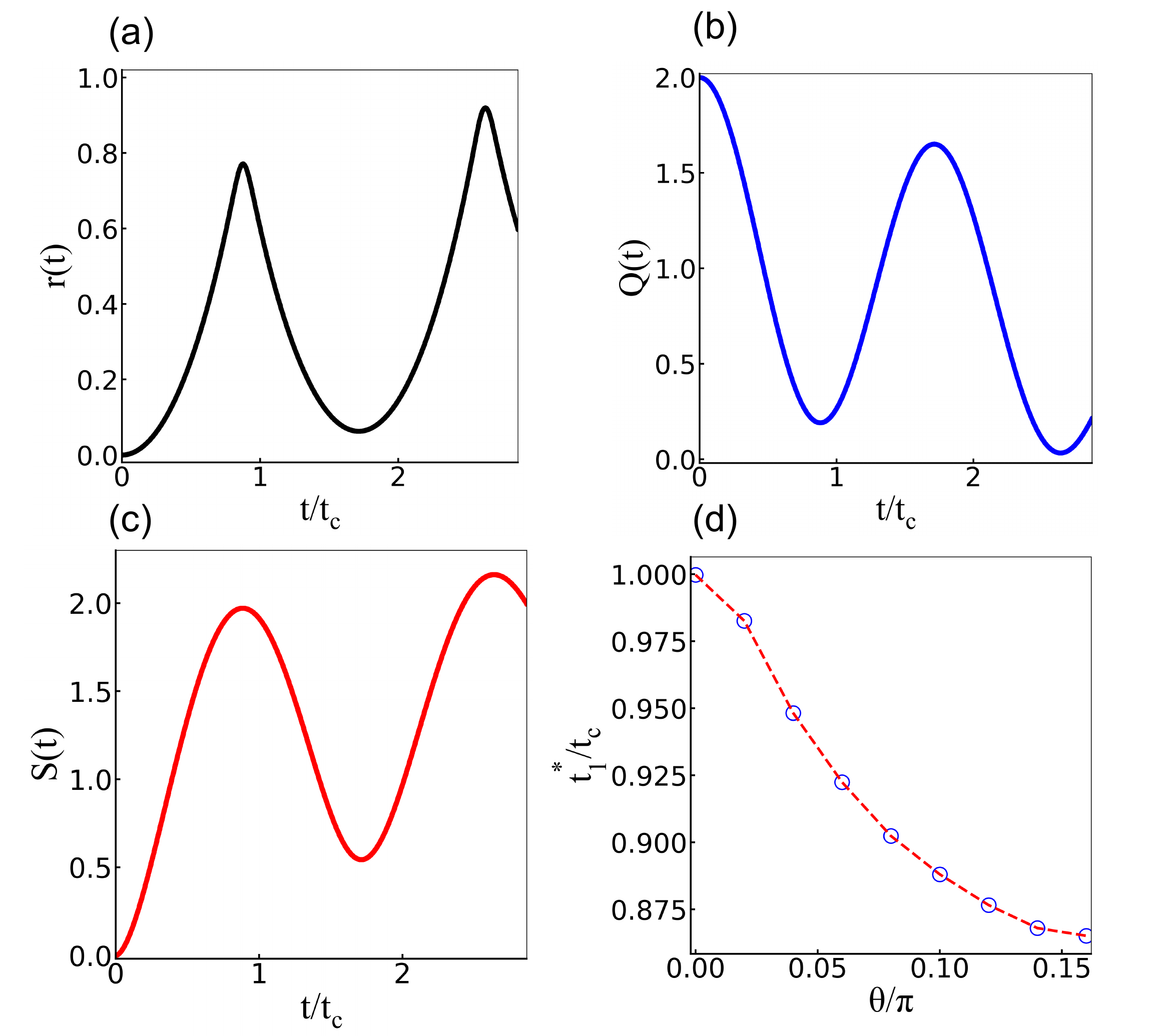}
\caption{(Color online) Time evolution of the return rate (a), dynamical order parameter (b), and entanglement entropy (c). The first critical time as a function of spatial chiral interaction $\theta$ (d). The figures are correspond to the quenches with $f_{0}/J = \infty$ (PM) $\rightarrow $ $f_{1}/J = 0.0$ (Potts FM), $\theta=0.12\pi, \phi=0, N=30$.}
\label{fig:potts2}
\end{figure}

We note that the asymmetry in the Hamiltonian has important consequences: the spatial chirality ($\theta \ne 0$) induces incommensurate floating phases with respect to the periodicity of the underlying lattice. To study whether chirality will affect the DQPT of the system, we numerically investigate the quantum chiral Potts chain with $\theta=0.12\pi$, quenches from the PM to the Potts FM phase. On the one hand, we start from the PM state, which is obtained from initial Hamiltonian $H_{0}$ with $f_{0} = 1.0, J = 0.0, f_{0} /J = \infty$, quenched to a final state with parameter$ f_{1} = 0.0, J = 1.0,f_{ 1}/J = 0.0$, located in Potts FM order phase. As shown in Fig~\ref{fig:potts2}(a), we find that the return rate exhibits a series of non-analytic behaviors with time, implying that the introduction of chirality is stable for observing DQPT. On the other hand, we also calculated the dynamical order parameter and entanglement entropy, as shown in Fig~\ref{fig:potts2}(b) and (c), and found that the valley (peak) of the dynamical order parameter (entanglement entropy) corresponds to the peak of the return rate, demonstrating that they share the same periodicity of the DQPT. Moreover, as shown in Fig~\ref{fig:potts2}(d), we found that the first critical time $t^{*}_{1}$ for the DQPT decreases with the increase of the chiral interaction (see the Appendix~\ref{sec:A12} for the calculation of the DQPT of another chiral interaction). This means that the increase in chiral interaction will make DQPT easier to occur,  we called "chirality-related dynamical phase transition". This adjustment allows for the observation of DQPTs in a shorter time, thereby reducing the experimental detection complexity associated with these transitions. In section ~\ref{sec:chiral} C, we will give a simple physical argument to understand this phenomenon. Finally, we also calculated the DQPT from the Potts FM to the PM phase with the temporal chiral interaction $\phi$. The results are shown in Appendix~\ref{sec:A2}.

\section{DYNAMICAL SCALING FOR CHIRAL CLOCK POTTS CHAIN}
\label{sec:chiral}
\subsection{Fidelity susceptibility and quantum critical point}
The system undergoes a continuous phase transition from an ordered to a disordered phase when tuning the external field $f$, at which the structure of the ground state wave function changes significantly. The quantum ground-state fidelity $F(f,f + \delta f)$, defined as the wave function overlap of two neighboring ground states with respect to an external field $f$, and its value is almost zero near quantum critical point $f_{c}^{*}$.
 In practice, the more convenient quantity to characterize quantum phase transitions is the fidelity susceptibility, defined by the leading term of fidelity~\cite{Gu_2014,gu2010fidelity,gu2009pre,Gu_2009,You_2015,yu2023quantum},

\begin{equation}
\begin{split}
\label{E9}
\chi_{F}(f)={\rm{lim}}_{\delta f \rightarrow 0}\frac{2(1-F(f,f+\delta f))}{(\delta f)^{2}}.
\end{split}
\end{equation} 

The fidelity susceptibility is a geometric property of quantum states in the realm of quantum information that offers a distinct advantage in that it requires no a priori knowledge of order parameters or symmetry-breaking. It has been applied to detect a wide range of quantum phase transitions~\cite{yu2014fidelity,sun2015prb,konig2016prb,yu2022prb_fs,sun2020prb,Tu2023generalpropertiesof} induced by a sudden change in the structure of the wave function. Experimental detection of quantum phase transitions using fidelity susceptibility can be achieved via neutron scattering or angle-resolved photoemission spectroscopy (ARPES) techniques~\cite{Gu_2014}. Here, we employ the fidelity susceptibility~\cite{albuquerque2010prb} to identify the critical point in the quantum Potts chain with chiral interaction and perform dynamical scaling in the obtained quantum critical point. 

\begin{figure*}[tb]
\includegraphics[width=0.8\textwidth]{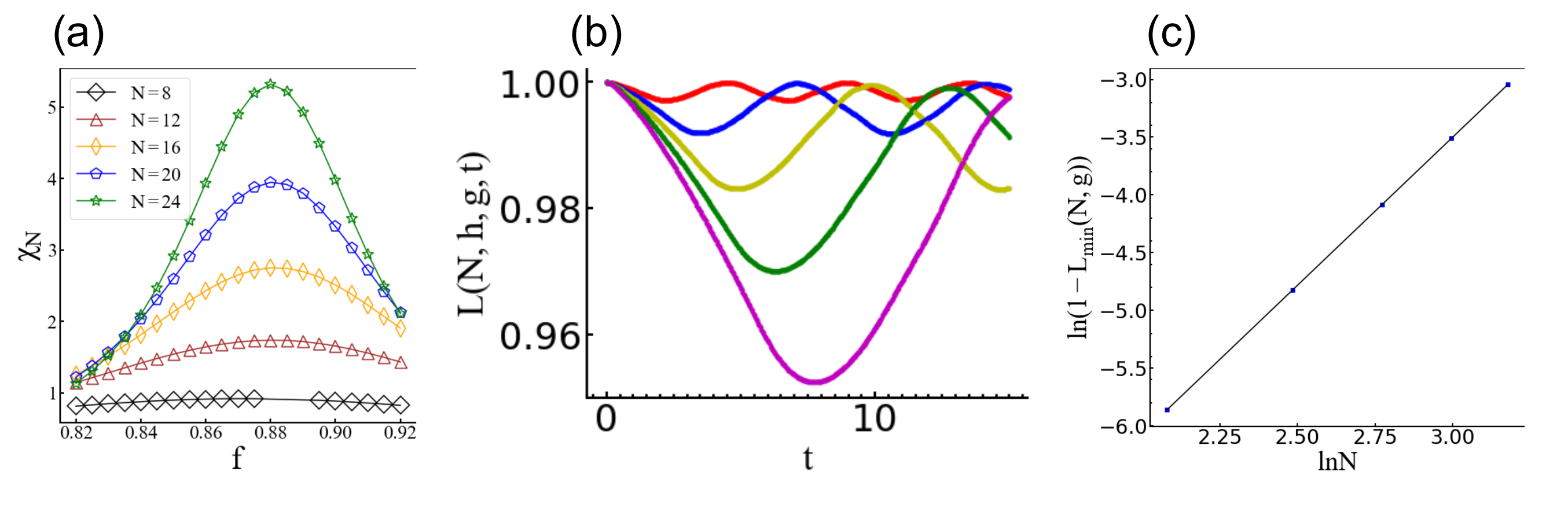}
\caption{(Color online) (a) Fidelity susceptibility per site $\chi_{N}$ of the quantum Potts chain with chiral interaction for $\theta=0.12\pi$ and $N=8,12,16,20,24$ sites as a function of external transverse field $f$; symbols denote finite-size DMRG results. (b) The Loschmidt echo $L(N,f,g,t)$ at the peak position $f$ of $\chi_{N}$ in (a) with $g=0.01$ as a function of time $t$ for lattice sizes $N=8$(red),$12$(blue),$16$(yellow),$20$(green),$24$(purple) (from top to bottom along the first minima). (c) Finite-size scaling of $1-L_{min}(N,g)$ obtained from (b) as a function of lattice sizes $N$ where the black square symbols are numerical values, and the block solid line denotes the fitting curve. The correlation length critical exponent $\nu=0.780 \pm 0.001$ is obtained from the fitting curve. }
\label{fig:fs}
\end{figure*}

Figure~\ref{fig:fs}(a) illustrates the scaling behavior of fidelity susceptibility per site $\chi_{N} = \chi_{F}/N$ as a function of system size $N$ for $\theta=0.12\pi$ in the quantum Potts chain with chiral interaction. As the system size increases from $N=8$ to $24$, we observe that the peak position of the fidelity susceptibility curve gradually approaches the exact critical point $f_{c}^{*}$. Our results indicate that the peak of fidelity susceptibility converges at $N=24$ (see Appendix~\ref{sec:A1} for the finite-size effect of DQPT), providing an effective means of obtaining the quantum critical point. (see Appendix~\ref{sec:A4} for details on the calculation of fidelity susceptibility for another chiral interaction.)

\subsection{Dynamical scaling law for Loschmidt echo}
As a next step, we aim to examine the dynamical property of the system following a small quench and carry out dynamical scaling of Loschmidt echo to obtain critical exponents in the non-conformal chiral universality class. Recent studies~\cite{hwang2019universality} have demonstrated that the decay of the Loschmidt echo can be enhanced by the equilibrium quantum criticality. The first minimum of the Loschmidt echo at the $t_{1}^{*}$  can be scale as:  
\begin{equation}
\label{E10}
1-L_{min}(N,g) \propto g^{2} N^{2/\nu},
\end{equation}
at the equilibrium chiral transition point. Here $\nu$ is the correlation length exponent, and $g$ is the small constant step define as: $g=f_{1}-f_{0}$ with $f_{0}$ and $f_{1}$ are external field before and after quench protocols. The exact definition of the first minimum of Loschmidt echo $L_{min}(N,g) = min_{t}|\langle \psi_{0}(f_{0}) |e^{-iH_{f}t}|\psi_{0}(f_{0})\rangle|^{2}$. The dynamical scaling law in Eq.~\ref{E10} that governs the critically enhanced decay behavior of the Loschmidt echo with respect to $N$ can be utilized to extract the correlation length exponent $\nu$. In order to perform the scaling law in Eq.\ref{E10} for the Loschmidt echo $L_{min}(N,g)$, it should be computed at or close to the equilibrium critical point $f_{c}^{*}$, which is obtained from fidelity susceptibility.

To this end, we first obtain the ground state $\psi_{0}$ from Eq.~\ref{E1} at the external field $f_{0}$, and then compute the Loschmidt echo from Eq.~\ref{E2} by quenching the chiral clock Potts chain from the initial $f_{0}$ to a final $f_{1}$ with a small constant step $g=0.01$. The time-evolved wave function $\psi(t)$ is obtained from the TDVP with the step $dt=0.02$ under periodic boundary conditions, where we set $J=1$ during the numerical simulations. We perform numerical simulations upon a quench with the TDVP from this critical point $f_{0}$ to $f_{1}=f_{0} + g$ for $N=8,12,16,20,24$ sites. The results of the Loschmidt echo $L(N,h,g,t)$ for $\theta=0.12\pi$ are shown in Fig.~\ref{fig:fs}(b) exhibit a decay and revival dynamics. The first minimum of the Loschmidt echoes $L_{min}(N,g)$ are plotted in Fig.~\ref{fig:fs}(c) with respect to the lattice size $N$. According to the scaling law in Eq.~\ref{E10}, we obtain the correlation length exponent $\nu=0.780 \pm 0.001$ for non-conformal chiral transition. Details on the dynamical scaling of Loschmidt echo for another chiral interaction can be found in Appendix.~\ref{sec:A3}, and the results of all $\theta$ are summarized in Table.~\ref{tab:exponents}. The results exhibit correlation length exponent $\nu$ decreases with increasing chiral interaction $\theta$, which is consistent with previous literatures~\cite{samjdar2018pra,yu2022prb_fs,yu2023quantum}. Moreover, we want to remark that when $\theta=\pi$, the model is simplified to a quantum antiferromagnetic Potts chain. Under this condition, the KT-like phase transition from the massive trivial phase to the massless phase will occur. The critical point is located at $f/J \approx 0.2$~\cite{verresen2023quantum,dai2017prb} (Due to the symmetry of the three-state chiral Potts model, $\theta=\theta+\pi/3$, $\theta=\pi$ is equivalent to $\pi/3$, therefore, the phase transition from massive trivial phase to massless incommensurate phase occurs~\cite{zhuang2015prb}). In the context of the BKT transition, the correlation length exhibits divergence characterized by $\xi \propto exp(C f^{-\nu}) \propto exp(C/\sqrt{f})$, where the "correlation length exponent" $\nu$ is known to be 0.5~\cite{chepiga2019prl,zuo2021prb}. However, a comprehensive understanding of DQPTs resulting from a quench across a BKT transition, as well as the associated dynamical scaling laws, remains an open question. We defer this intriguing topic to future work.

\begin{ruledtabular}
\begin{table}[tb]
\caption{Critical exponents of the Potts chain with different chiral interaction $\theta$. Critical exponents are obtained by dynamical scaling for the Loschmidt echoes.}
\label{tab:exponents}
\begin{tabular}{cc}
$\theta$	& $\nu$ \\ \hline
0.0$\pi$	& 0.838(4)					\\
0.02$\pi$	& 0.832(4)				\\
0.04$\pi$	& 0.829(3)								\\
0.06$\pi$	& 0.824(3)										\\
0.08$\pi$   & 0.813(1)                                          \\
0.10$\pi$   & 0.7978(5)                                         \\
0.12$\pi$   & 0.780(1)                                          \\
0.14$\pi$   & 0.754(2)                                        \\
0.16$\pi$   & 0.724(3)                                     \\
\end{tabular}
\end{table}
\end{ruledtabular}

\subsection{Discussion}
In this section, we present a brief discussion on the physical reason for the emergence of chirality-related dynamical phase transition. Firstly, the introduction of the spatial chiral interaction $\theta$ can be considered as an equivalent of introducing power-law long-range interactions. To be precise, when the spatial chiral interaction is increased (i.e., larger $\theta$), the interaction becomes more long-range. Recent studies~\cite{samjdar2018pra} on Rydberg atom arrays have demonstrated that the van der Waals long-range interaction has a similar effect on the critical behavior as the chiral interaction. In particular, the longer the power-law interaction, the smaller the critical exponent of the correlation length $\nu$, which is consistent with the results obtained by dynamical scaling (see Table~\ref{tab:exponents}). This is because the long-range interactions enhance the in-equivalence between the two types of domain walls, leading to a faster deviation from the Potts exponent~\cite{yu2022prb_fs}.

On the other hand, previous literature~\cite{ifmmode2018prl,jad2017prb,jad2017pre,jad2017prb,jad2018prb,jad2018prl,jad2020prr,halimeh2019dynamical,jad2021prb,jad2021prb_b} reports have studied the dynamical phase transitions in the transverse field Ising model with power-law long-range interactions $1/r^{\alpha}$. According to Fig.1 of Ref.~\cite{ifmmode2018prl}, when $\alpha \textless 2.0$  (where the physics is essentially analogous to that of a short-range Ising model for large $\alpha$), we observed that a smaller power-law long-range interaction exponent $\alpha$ implies a stronger long-range interaction, and consequently, effective enhance chirality, as discussed previously. This results in a lower critical transverse field required for the appearance of kinks in the Loschmidt echo under identical quenching conditions, which means that DQPT will occur earlier, that is, the first critical time will decrease (advance).

By combining the above two aspects, we provide a simple physical explanation for chirality-related dynamical phase transition: The effect of spatial chiral interaction is akin to that of long-range interaction, and the introduction of long-range interaction enhances the propensity for DQPT in the system. Therefore, spatial chiral interaction can effectively advance dynamical phase transition. Finally, we observe that the inclusion of temporal chiral interaction leads to different results (see Appendix~\ref{sec:A2}), which merits further detailed investigation in the future.

\section{CONCLUSION AND OUTLOOK}
\label{sec:con}
To summarize, we investigate the quench dynamics in the $\mathbb{Z}_{3}$ symmetric spin chain with chiral interaction. To establish a baseline, we first consider the standard nearest-neighbor quantum three-state Potts chain and derived analytical results of the Loschmidt echo for the quench from the PM to the Potts FM phase. Our results show that the Loschmidt echo, dynamical order parameter, and entanglement entropy both exhibit DQPT signatures. We then investigate more general cases with a chiral interaction $\theta$ using TDVP algorithms. The results reveal that the introduction of the chiral interaction will advance the critical time of DQPT, which we refer to as "chirality-related dynamical phase transition". Additionally, we perform dynamical scaling for the Loschmidt echo and obtain the correlation length critical exponent $\nu$ under different chiral interactions. The numerical results indicate that as the chiral interaction increases, the correlation length exponent $\nu$ decreases, which has a similar effect of long-range interaction. Finally, we provide a simple physical argument to understand "chirality-related dynamical phase transition."

Future work may explore the physical reason for the dynamical phase transition due to temporal chiral interaction and investigate the fate of quench dynamics in two-dimensional systems with different types of quantum critical points. Our work may provide new insights into many-body physics out of equilibrium with chiral interaction.

\begin{acknowledgments}
X.-J. Yu thank G. Sun, S.Yang, and C.-X. Li for helpful discussions. Numerical simulations were carried out with the ITensor package~\cite{matthew2022sci}. X.-J.Yu acknowledges support from the start-up grant of Fuzhou University.
\end{acknowledgments}

\bibliography{DQPT_chiral_clock_arxiv}

\newpage
\onecolumngrid

\appendix

\section{FINITE SIZE AND RIME STEP EFFECT OF DQPT}
\label{sec:A1}
\begin{figure}[tb]
\includegraphics[width=0.7\textwidth]{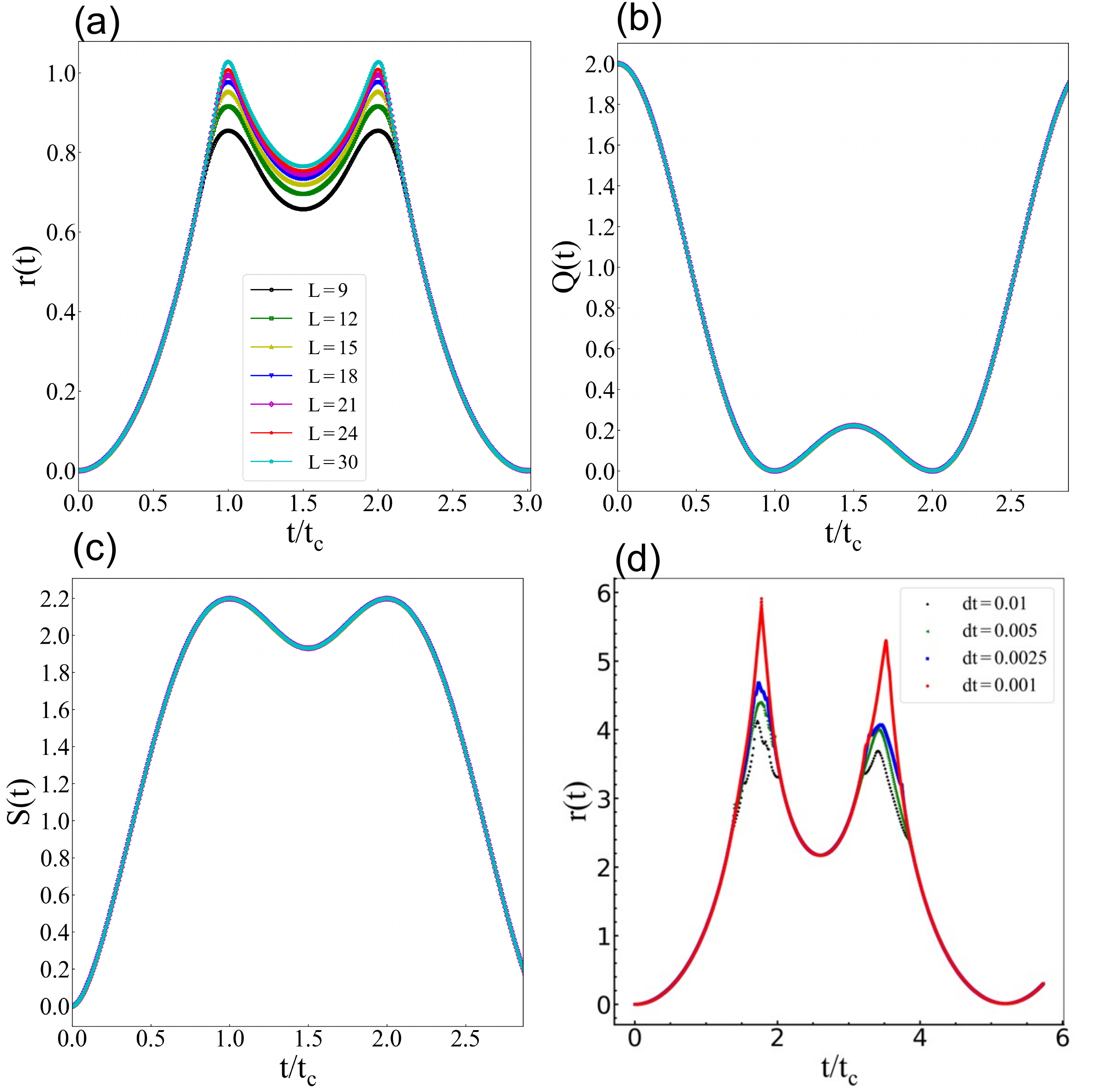}
\caption{(Color online) (a) The return rate as a function of time for system size $N$ from 9 to 30 sites across the critical point.  (b) Dynamical order parameter, and (c) entanglement entropy. It is clear that the return rate $r(t)$ of the largest four system sizes is quite good with only a slight deviation of the smallest two sizes. All the figures are correspond to the from PM state, quench to final Potts FM state. (d) The return rate as a function of time for different time step $dt$ from 0.01 to 0.001 across the critical point for $\phi=0.12\pi,\theta=0.0$, $N=30$. The figure is correspond to the quenches from Potts FM state, quench to final PM phase. }
\label{fig:fss}
\end{figure}

In this Appendix, we investigate the effect of the finite system sizes on the extraction of the dynamical properties, namely, the associated return rate $r(t)$, dynamical order parameter $Q(t)$, and entanglement entropy $S(t)$. To this end, we perform TDVP calculation for different system size $N$ from 9 to 30 sites with periodic boundary conditions and consider a quench originating in the PM phase with $\phi=\theta=0,f_{0}=1.0, J=0.0, f_{0}/J=\infty$, and subsequently quench to the Potts FM phase with $f_{1}=0.0, J=1.0, f_{1}/J=0.0$, as shown in Fig. ~\ref{fig:fss}. It is clear that the return rate $r(t)$ of the largest four system sizes is quite good with only a slight deviation of the smallest two sizes, which indicates that the system size $N=30$ is sufficient to detect the DQPT, and we notice that dynamical order parameter and entanglement entropy does not appear as a shape peak, but exhibits the time evolution of the same period as the return rate. It does not change with system size. Therefore, we can expect that the figures in the main text are reliable to represent the results in the thermodynamic limit. In addition, we also calculated the return rate under different time steps $dt=0.01,0.005,0.0025,0.001$, and the results showed that the result of $dt=0.001$ basically converged, as shown in Fig.~\ref{fig:fss} (d).

\section{DYNAMICAL PHASE TRANSITION EVOLVES WITH CHIRAL INTERACTION}
\label{sec:A12}
In this section, we provide additional data to show dynamical phase transition evolves with chiral interaction. 

\begin{figure}[tb]
\includegraphics[width=0.5\textwidth]{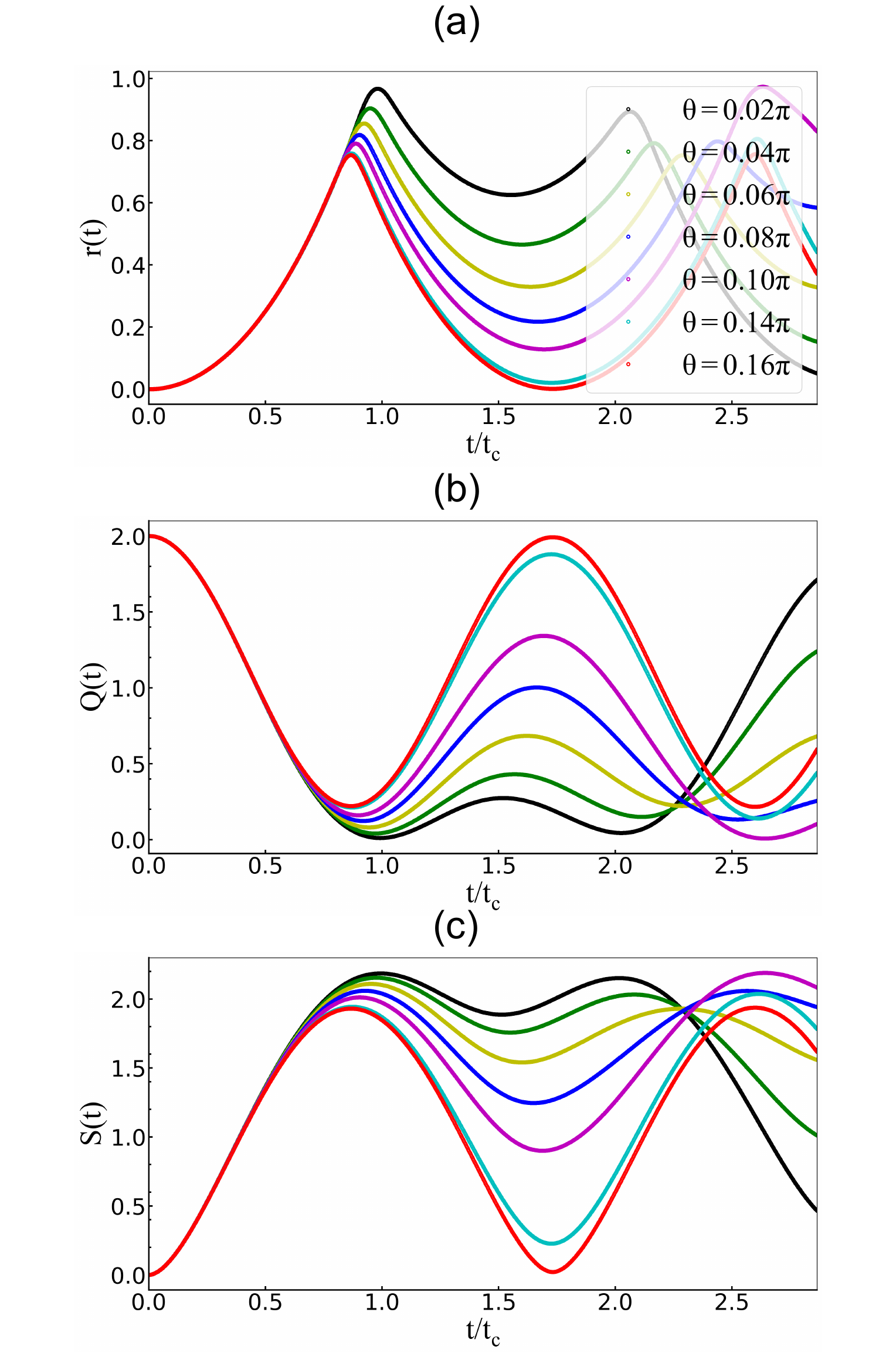}
\caption{(Color online) Time evolution of the return rate (a), dynamical order parameter (b), and entanglement entropy (c) of the quantum Potts chain with spatial chiral interaction ($\phi=0.0$) for $\theta=0.02\pi,0.04\pi,0.06\pi,0.08\pi,0.1\pi,0.14\pi,0.16\pi$, and $N=30$.  All the figures are correspond to the quenches with $f_{0}/J = \infty$ (PM)   $\rightarrow $ $f_{1}/J = 0.0$ (Potts FM).}
\label{fig:theta3}
\end{figure}

As the same in the main text, return rate  $r(t)$ of the Potts chain with chiral interaction for (a) $\theta=0.0\pi$,(b) $\theta=0.02\pi$,(c) $\theta=0.04\pi$,(d) $\theta=0.06\pi$, (e) $\theta=0.08\pi$, (f) $\theta=0.1\pi$, (g) $\theta=0.14\pi$, (h) $\theta=0.16\pi$, and $N=8,12,16,20,24$ sites as a function of time $t$, are shown in the Fig.~\ref{fig:theta3}. We found that no matter in return rate, dynamical order parameter, and entanglement entropy, the time when kink (that is, DQPT) appears for the first time decreases with the increase of $\theta$, which is a chirality-related dynamical phase transition in the main text.

\section{WITH TEMPORAL CHIRAL INTERACTION: QUENCH FROM POTTS TO PM}
\label{sec:A2}
\begin{figure}[tb]
\includegraphics[width=0.5\textwidth]{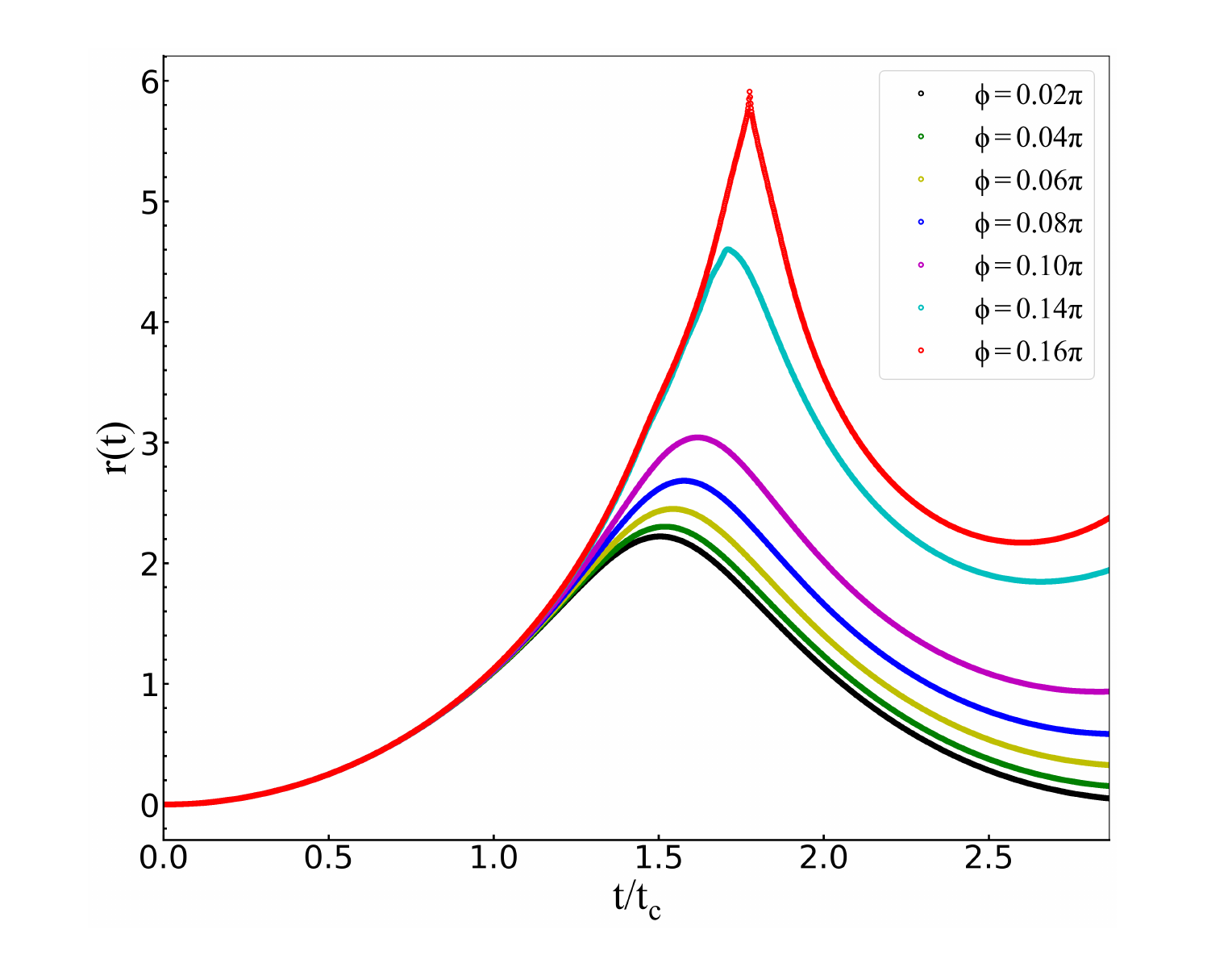}
\caption{(Color online) Time evolution of the return rate of the quantum Potts chain with temporal chiral interaction ($\theta=0.0$) for $\phi=0.02\pi,0.04\pi,0.06\pi,0.08\pi,0.1\pi,0.14\pi,0.16\pi$, and $N=30$.  The figure are correspond to the quenches with $f_{0}/J = 0.0$ (Potts FM)   $\rightarrow $ $f_{1}/J = \infty$ (PM).}
\label{fig:phi3}
\end{figure}

\begin{figure}[tb]
\includegraphics[width=0.8\textwidth]{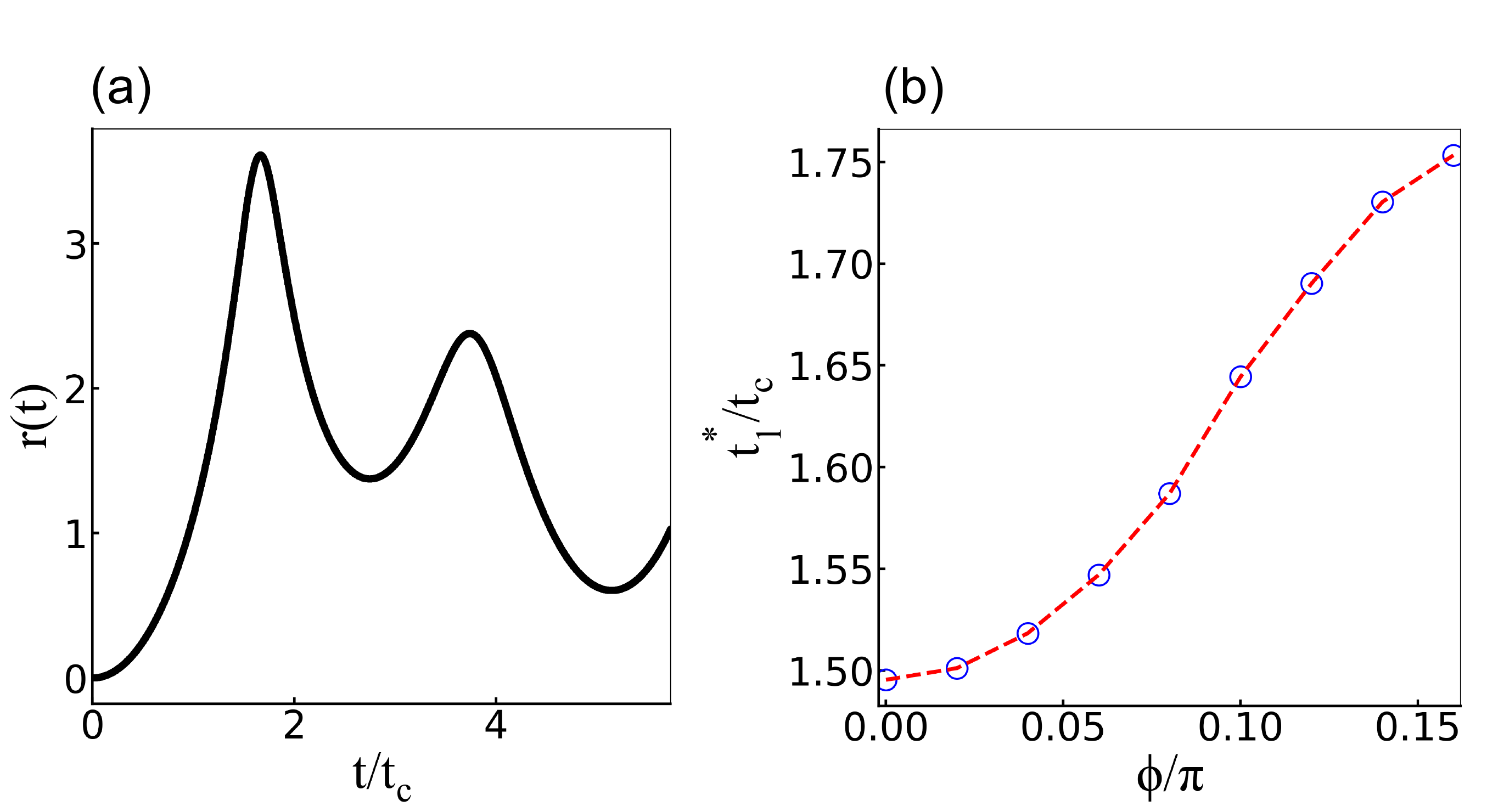}
\caption{(Color online) (a) Time evolution of the return rate. The figures are correspond to the quenches with $f_{0}/J = 0.0$ (Potts FM)   $\rightarrow $ $f_{1}/J = \infty$(PM), $\theta=0.0, \phi=0.12\pi, N=30$. (b) The first critical time is a function of temporal chiral interaction phase $\phi$.}
\label{fig:chiral1}
\end{figure}

We briefly discuss the DQPT of the Potts model with temporal chiral interaction. To this end, we start from Potts FM ($f_{0}=0.0, J=1.0,f_{0}/J=0.0$), quench to PM state ($f_{1}=1.0, J=0.0,f_{1}/J=\infty$) with temporal chiral interaction phase $\phi$. On the one hand, the return rate of the Potts chain with chiral interaction for (a) $\phi=0.0\pi$,(b) $\phi=0.02\pi$,(c) $\phi=0.04\pi$,(d) $\phi=0.06\pi$, (e) $\phi=0.08\pi$, (f) $\phi=0.1\pi$, (g) $\phi=0.14\pi$, (h) $\phi=0.16\pi$, and $N=8,12,16,20,24$ sites as a function of time $t$, both exhibit singular behavior,  are shown in the Fig.~\ref{fig:phi3}. On the other hand, combined with the return rate, we give the first critical time $t_{1}^{*}$ as a function of temporal chiral interaction phase $\phi$ for $\phi=0.12\pi$,$N=30$, as shown in the Fig~\ref{fig:chiral1}. The results show that $t_{1}^{*}$ increases with the increase of $\phi$, in contrast to the spatial chiral interaction case. 

\section{DYNAMICAL SCALING FOR OTHER CHIRAL INTERACTION PHASES}
\label{sec:A3}
In this section, we provide additional data to show dynamical scaling for other chiral interaction phases. 

\begin{figure*}[tb]
\includegraphics[width=1.0\textwidth]{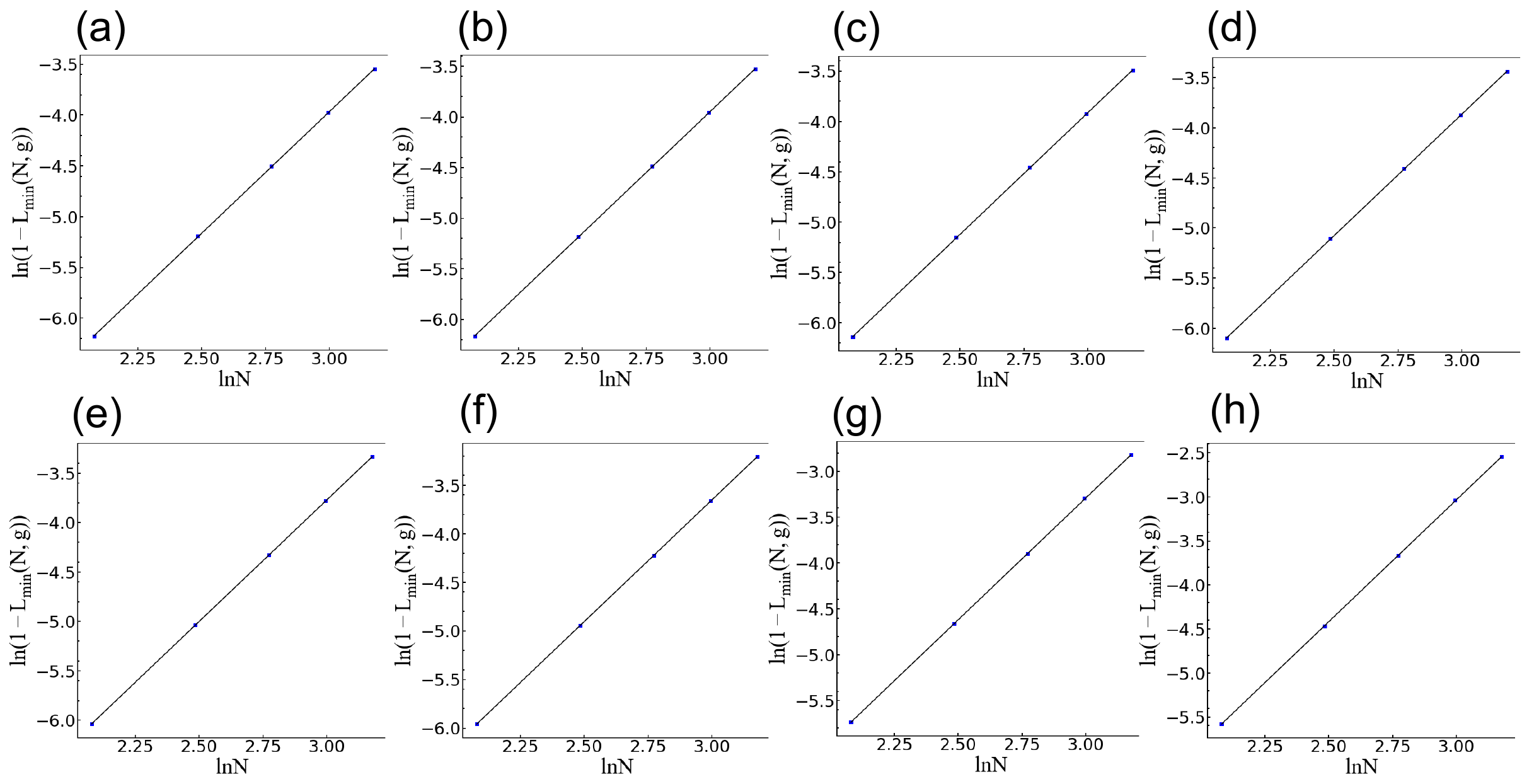}
\caption{(Color online) Finite-size scaling of $1-L_{min}(N,g)$  of the quantum Potts chain with spatial chiral interaction ($\phi=0.0$) for (a) $\theta=0.0\pi$ ,(b) $\theta=0.02\pi$,(c) $\theta=0.04\pi$,(d) $\theta=0.06\pi$, (e) $\theta=0.08\pi$, (f) $\theta=0.1\pi$, (g) $\theta=0.14\pi$, (h) $\theta=0.16\pi$, and $N=8,12,16,20,24$ sites as a function of lattice sizes $N$; symbols denote finite-size DMRG results.}
\label{fig:appC}
\end{figure*}

As the same in the main text, the dynamical scaling for $1-L_{min}(N,g)$ in the Potts chain with chiral interaction for (a) $\theta=0.0\pi$,(b) $\theta=0.02\pi$,(c) $\theta=0.04\pi$,(d) $\theta=0.06\pi$, (e) $\theta=0.08\pi$, (f) $\theta=0.1\pi$, (g) $\theta=0.14\pi$, (h) $\theta=0.16\pi$, and $N=8,12,16,20,24$ sites as a function of lattice sizes $N$, are shown in the Fig.~\ref{fig:appC}. The extrapolated correlation length exponents are summarized in Table~\ref{tab:exponents}, we found that with the enhancement of chiral interaction $\theta$, the correlation length exponent $\nu$ of the system gradually decreases, which is consistent with the effect of long-range interaction.

\section{FIDELITY SUSCEPTIBILITY FOR OTHER CHIRAL INTERACTION PHASES}
\label{sec:A4}
In this section, we provide additional data to show fidelity susceptibility for other chiral interaction phases. 

\begin{figure*}[tb]
\includegraphics[width=1.0\textwidth]{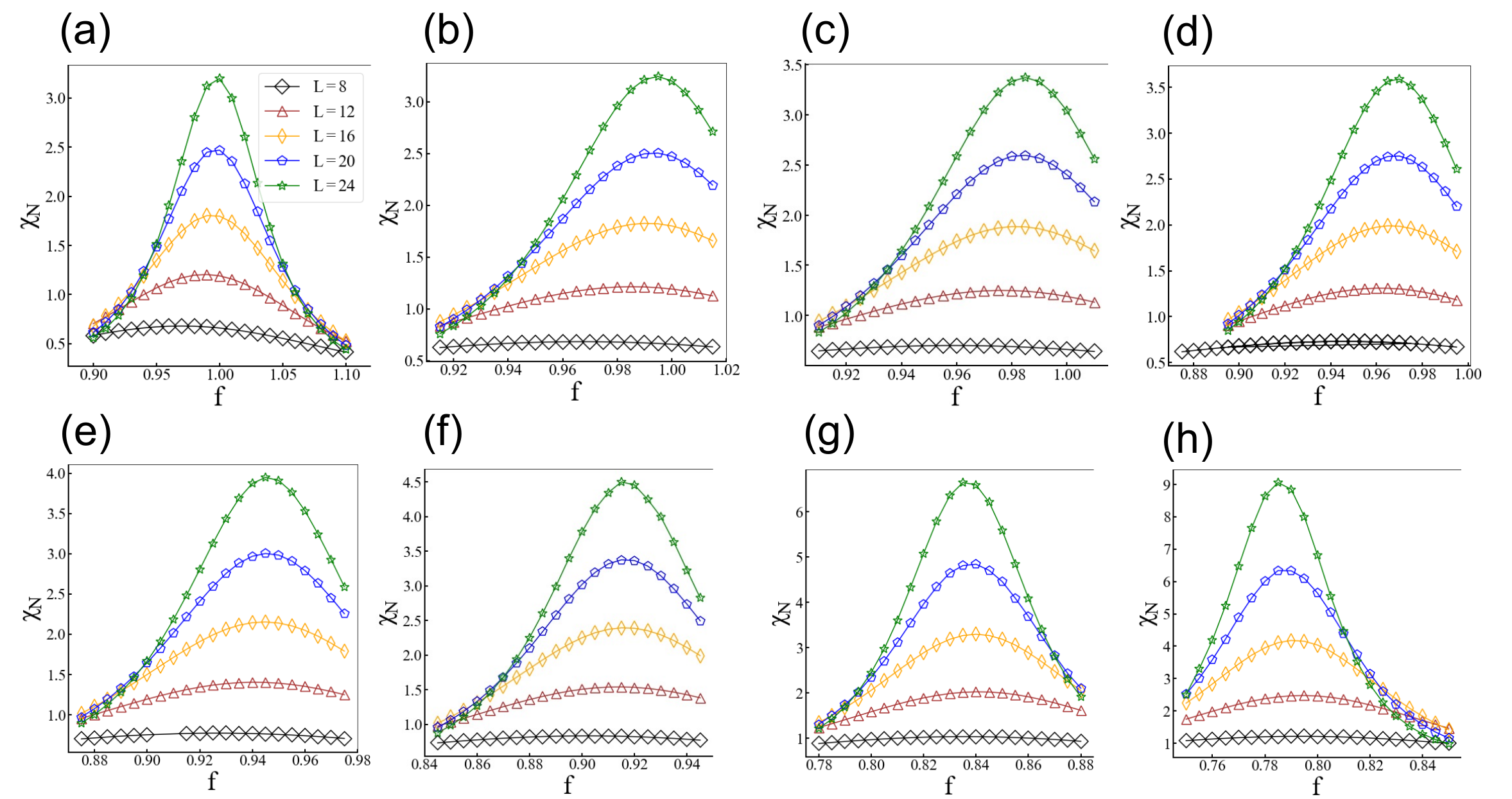}
\caption{(Color online)  Fidelity susceptibility per site $\chi_{N}$ of the quantum Potts chain with spatial chiral interaction ($\phi=0.0$) for (a) $\theta=0.0\pi$ ,(b) $\theta=0.02\pi$,(c) $\theta=0.04\pi$,(d) $\theta=0.06\pi$, (e) $\theta=0.08\pi$, (f) $\theta=0.1\pi$, (g) $\theta=0.14\pi$, (h) $\theta=0.16\pi$, and $N=8,12,16,20,24$ sites as a function of external transverse field $f$ ; symbols denote finite-size DMRG results.}
\label{fig:appD}
\end{figure*}

As the same in the main text, on the one hand, fidelity susceptibility per site $\chi_{N}=\chi_{F}/N$ of the Potts chain with chiral interaction for (a) $\theta=0.0\pi$,(b) $\theta=0.02\pi$,(c) $\theta=0.04\pi$,(d) $\theta=0.06\pi$, (e) $\theta=0.08\pi$, (f) $\theta=0.1\pi$, (g) $\theta=0.14\pi$, (h) $\theta=0.16\pi$, and $N=8,12,16,20,24$ sites as a function of the external transverse field $f$, are shown in the Fig.~\ref{fig:appD}. We find that quantum critical points are shifted to lower values of $f$ as chiral interaction increases.

\section{LOSCHMIDT ECHO AT THE PEAK POSITION FOR OTHER CHIRAL INTERACTION PHASE}
\label{sec:A5}
In this section, we provide additional data to show Loschmidt echo at the peak position for other chiral interaction phases. 

\begin{figure*}[tb]
\includegraphics[width=1.0\textwidth]{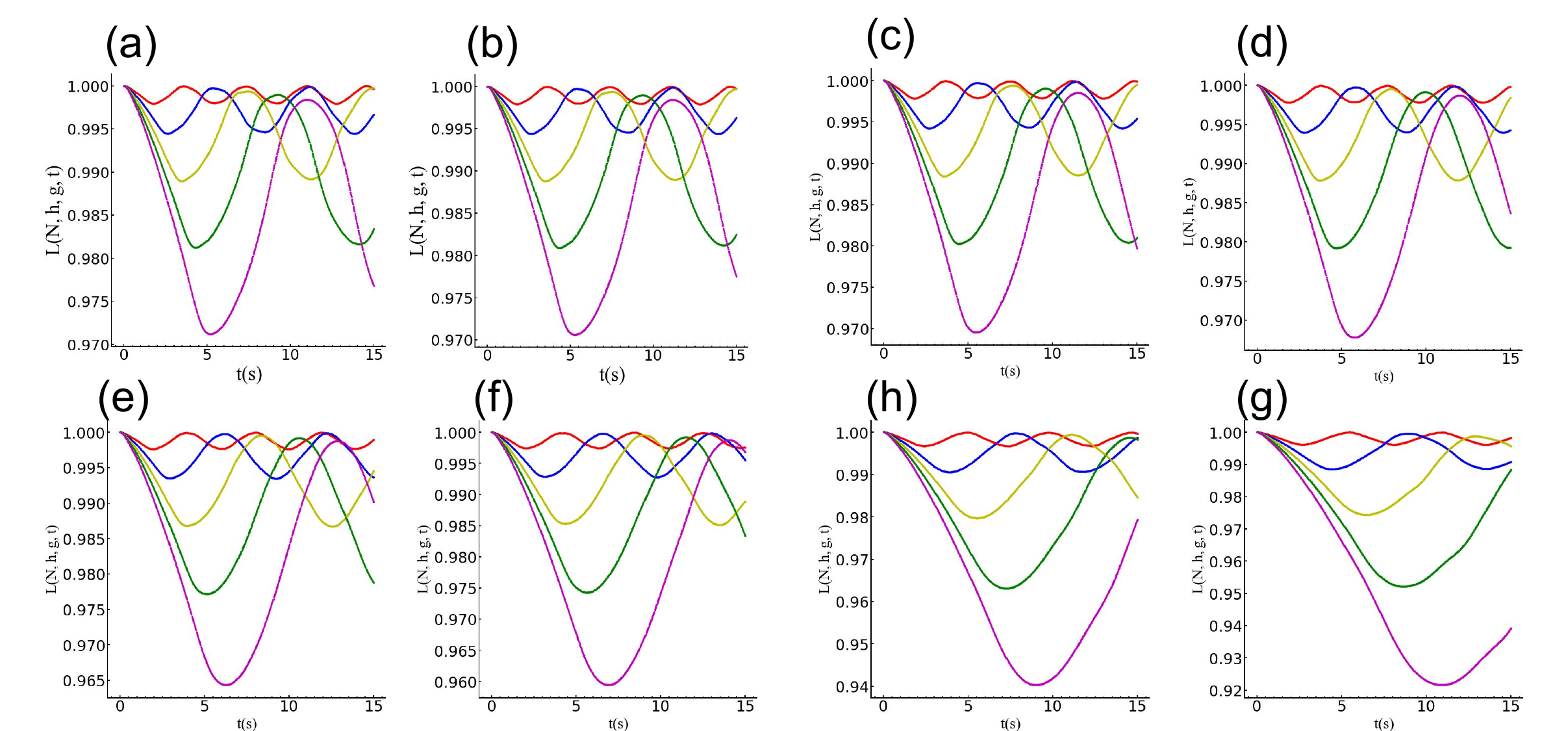}
\caption{(Color online) The Loschmidt echo $L(N,f,g,t)$ at the peak position $f$ of $\chi_{L}$ with $g=0.01$ of the quantum Potts chain with spatial chiral interaction ($\phi=0.0$) for (a) $\theta=0.0\pi$, (b) $\theta=0.02\pi$, (c) $\theta=0.04\pi$, (d) $\theta=0.06\pi$, (e) $\theta=0.08\pi$, (f) $\theta=0.1\pi$, (g) $\theta=0.14\pi$, (h) $\theta=0.16\pi$, and $N=8$ (red), 12 (blue), 16 (yellow), 20 (green), 24 (purple) sites as a function of time $t$; symbols denote finite-size TDVP results. }
\label{fig:appE}
\end{figure*}

As the same in the main text, the Loschmidt echo at the peak position of the Potts chain with chiral interaction for (a) $\theta=0.0\pi$, (b) $\theta=0.02\pi$, (c) $\theta=0.04\pi$, (d) $\theta=0.06\pi$, (e) $\theta=0.08\pi$, (f) $\theta=0.1\pi$, (g) $\theta=0.14\pi$, (h) $\theta=0.16\pi$, and $N=8,12,16,20,24$ sites as a function of time $t$, are shown in the Fig.~\ref{fig:appE}. We found that the time to the first appearance of the kink increases with increasing chiral interaction.

\section{QUENCH FROM FM POTTS ORDER TO PM PHASE WITH SPATIAL CHIRAL INTERACTION AND BOND DIMENSION DEPENDENT}
\label{sec:A6}
\begin{figure}[tb]
\includegraphics[width=0.8\textwidth]{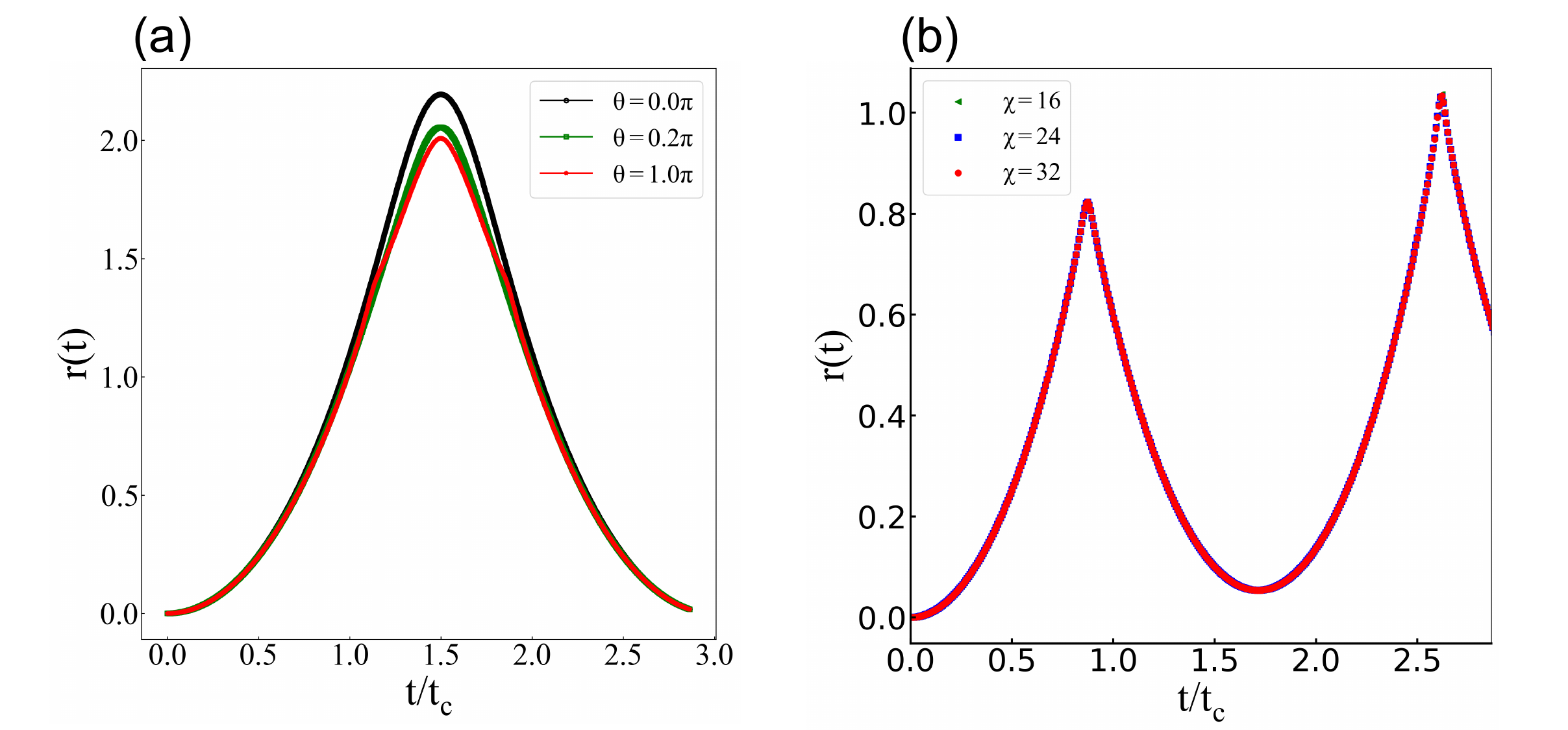}
\caption{(Color online) The return rate as a function of time for chiral interaction $\theta$ (a) for $0.0\pi$, $0.2\pi$ and $1.0\pi$ and (b) for different bond dimension $\chi=16,24,32$ for $\theta=0.12\pi$ across the critical point. Starting from Potts FM phase, quench to final PM phase for $N=30$. }
\label{fig:appG}
\end{figure}

In this appendix, we conduct TDVP simulations across various chiral interaction strengths ($\theta=0.0\pi, 0.2\pi$, and $1.0\pi$), while utilizing periodic boundary conditions. To be more precise, we consider a quench starting from the Potts FM phase with $\phi=0$, $f_{0}=0.0$, $J=1.0$, and $f_{0}/J=0.0$, and subsequently quenching to the PM phase with $f_{1}=1.0$, $J=0.0$, and $f_{1}/J=\infty$, as depicted in Fig.~\ref{fig:appG}(a). Notably, when $\theta=0$, as previously established, there is no occurrence of DQPT in this case\cite{karrasch2017prb}. As we introduce the chiral interaction, we observe a gradual smoothing of the peak value of the return rate, and yet the system still does not undergo DQPT.

Furthermore, we vary the MPS bond dimension ($\chi$) to investigate its impact on the return rate, as the return rate is directly related to the observation of DQPT, as shown in Fig.~\ref{fig:appG} (b). The results reveal that the return rate converges when $\chi$ exceeds $32$. Therefore, utilizing $\chi=60$ in the main text for the dynamical calculations is sufficient to ensure result convergence.

\end{document}